\documentstyle[epsf]{sup}

\makeatletter
\gdef\@journal{\normalsize{\it submitted to Superlattices and
Microstructures}}
\def\ps@titlepage{\let\@mkboth\@gobbletwo
 \def\@oddhead{\hss\vbox{\hsize=\textwidth \hbox to \textwidth
		{\strut\small\rm\@journal\hfill}}}
 \def\@evenhead{\hss\vbox{\hsize=\textwidth \hbox to \textwidth
		{\strut\small\rm\@journal\hfill}
	       \vskip 2.5pt \vskip \arrayrulewidth}}
 \def\@oddfoot{\normalsize \strut}
 \def\sectionmark##1{}
 \def\subsectionmark##1{}
}
\def\endabstract{\par%
    \hspace*{\fill}%
    \endlist%
    \vspace{-8pt}%
    \ignorespaces%
    \rlap{\bottom@eightrule}%
    \vspace{-6pt}}
\makeatother
\vspace{-8pt}

\title[Mesoscopic Josephson Effect]%
{Mesoscopic Josephson Effect}
\author[Hermann Grabert and Gert-Ludwig Ingold]%
{Hermann Grabert\cr%
{\normalsize\it Fakult\"at f\"ur Physik, Albert-Ludwigs-Universit{\"a}t,
Hermann-Herder-Stra{\ss}e~3,}\cr\vspace{10pt}
{\normalsize\it D-79104 Freiburg, Germany}\cr
Gert-Ludwig Ingold\cr
{\normalsize\it Institut f\"ur Physik, Universit\"at Augsburg,
Universit\"atsstra{\ss}e~1,}\cr
{\normalsize\it D-86135 Augsburg, Germany}\cr
{\normalsize and}\cr
{\normalsize\it CEA, Service de Physique de l'Etat
Condens\'e, Centre d'Etudes de Saclay,}\cr
{\normalsize\it F-91191 Gif-sur-Yvette, France}
}

\pagerange{\pageref{firstpage}--\pageref{lastpage}}

\def\dd{{\rm d}}     % Differential dee     %
\def\ee{{\rm e}}     % Exponential e        %
\def\ii{{\rm i}}     % Imaginary i          %
\begin{document}
\label{firstpage}
\maketitle
\sloppy
\begin{center}
\received{(\today)}
\end{center}
%**********************************************************************%
%
\begin{abstract}
In the classical Josephson effect the phase difference across
the junction is well defined, and the supercurrent is reduced 
only weakly by phase diffusion. For mesoscopic junctions with 
small capacitance the phase undergoes large quantum fluctuations, 
and the current is also decreased by Coulomb blockade effects. We 
discuss the behavior of the current-voltage characteristics
in a large range of parameters comprising the phase diffusion 
regime with coherent Josephson current as well as the supercurrent 
peak due to incoherent Cooper pair tunneling in the Coulomb blockade 
regime.
\end{abstract}
%
%%%%%%%%%%%%%%%%%%%%%%%%%%%%%%%%%%%%%%%%%%%%%%%%%%%%%%%%%%%%%%%%%%%%%%%%%

\section{Introduction}
The Josephson effect beautifully embodies quantum tunneling and 
superconductivity \cite{barone,tinkham}. Josephson systems
have been a subject of fundamental research for decades, which 
have led to useful applications, e.g., in highly
sensitive magnetometry and metrology. Like in the Ginzburg--Landau
theory for bulk superconductivity, where the phase of the
order parameter is considered as a quasiclassical variable,
the phase difference across a Josephson junctions can frequently
be treated classically. However, in the last ten years or so,
new lithography and low-temperature techniques have allowed the
fabrication and measurement of small Josephson junctions affected
by the capacitive charging energy of single Cooper 
pairs \cite{schoe90,sct91}. A large charging energy will render
the Cooper pair number on either side of the junction classical
and thus cause large quantum fluctuations of the conjugate phase
variable. 

In this article we show how the classical Josephson effect,
that is a supercurrent $I$ flowing at vanishing voltage $V$,
gradually evolves via the classical phase diffusion regime into the
supercurrent peak in the Coulomb blockade regime where Cooper
pairs tunnel incoherently across the Josephson junction.
After a brief review of the classical Josephson effect in
sec.~\ref{sec:classicaljosephsoneffect} and phase diffusion in
sec.~\ref{sec:phasediffusion} we discuss Cooper pair tunneling
in the Coulomb blockade regime in sec.~\ref{sec:coulombblockade}.
The turnover between the classical and the quantum regimes
is discussed in sec.~\ref{sec:between}. Finally, 
sec.~\ref{sec:conclude} contains some concluding remarks.

\section{Classical Josephson Effect}
\label{sec:classicaljosephsoneffect}

\begin{figure}
\begin{center}
\leavevmode
\epsfxsize=0.58\textwidth
\epsfbox{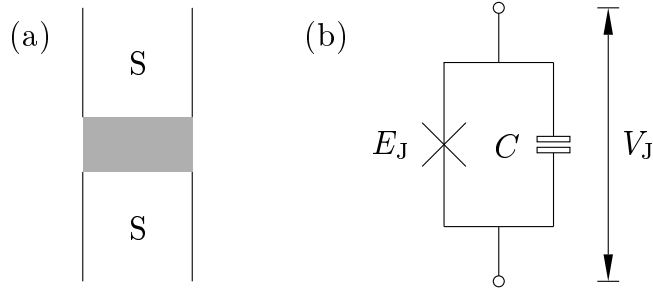}
\end{center}
\caption{(a) Schematic view of a Josephson junction where two
superconductors are separated by an insulating barrier. (b)
Equivalent circuit with Josephson coupling $E_{\rm J}$ and
capacitance $C$. The voltage across the junction is denoted by
$V_{\rm J}$.}
\label{fig:jj}
\end{figure}

A Josephson junction consists of two superconductors separated by a
thin insulating barrier as shown in fig.~\ref{fig:jj}a.
Its dynamics may be described in terms of the phase difference $\varphi$ 
of the two condensate wave functions on the left and right sides of 
the junction and the number $n$ of Cooper pairs on the capacitor formed 
by the junction. These operators $n$ and $\varphi$ obey the commutation 
relation $[n,\varphi]=-\ii$, and the junction Hamiltonian may be 
expressed as
\begin{equation}
H_{\rm J} = E_{\rm c}n^2 - E_{\rm J}\cos(\varphi).
\label{eq:hamil1}
\end{equation}
The first term describes the charging energy associated with the
capacitance $C$ of the tunnel junction where $E_{\rm c} = 2e^2/C$ 
is the charging energy corresponding to a single Cooper pair. The 
second term arises from the tunneling of Cooper pairs through the 
junction characterized by the Josephson coupling energy $E_{\rm J}$. 
These properties may be expressed in terms of the circuit shown
in fig.~\ref{fig:jj}b.

The Josephson relations link the phase to the voltage across the 
junction
\begin{equation}
V_{\rm J} = \frac{\hbar}{2e}\dot\varphi
\label{eq:josrelv}
\end{equation}
and to the supercurrent
\begin{equation}
I = I_{\rm c}\sin(\varphi).
\label{eq:josreli}
\end{equation}
The critical current $I_{\rm c}$ is related to the Josephson coupling 
energy by $I_{\rm c} = 2eE_{\rm J}/\hbar$. These relations allow for a
supercurrent flowing at zero voltage if $\varphi$ remains constant in 
time.

\section{Phase Diffusion in Josephson Junctions}
\label{sec:phasediffusion}
 
In the real world Josephson junctions are coupled to an electromagnetic
environment which may be described by an impedance $Z(\omega)$.  For 
simplicity, we will in the following mostly consider the case of an
ohmic resistor [$Z(\omega)=R$] but a more general case will be addressed
in sec.~\ref{sec:coulombblockade}. The resistor will give rise to 
Nyquist noise and therefore to a diffusive behavior of the phase 
difference $\varphi$.

\begin{figure}
\begin{center}
\leavevmode
\epsfxsize=0.35\textwidth
\epsfbox{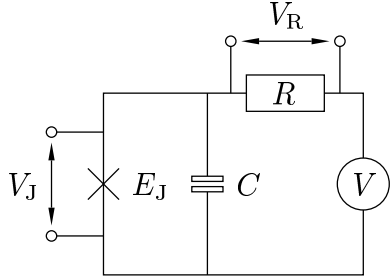}
\end{center}
\caption{Circuit modelling a real experimental set-up: a Josephson 
junction characterized by the Josephson energy $E_{\rm J}$ and 
capacitance $C$ is coupled to a voltage source $V$ via a 
resistor $R$. The voltage drops across the junction and the resistor are
$V_{\rm J}$ and $V_{\rm R}$, respectively.}
\label{fig:circuit}
\end{figure}

To be specific, we consider the circuit shown in fig.~\ref{fig:circuit}
where the Josephson junction is coupled to the voltage source via an
ohmic resistor. The voltages across the junction and the resistor are 
denoted be $V_{\rm J}$ and $V_{\rm R}$, respectively. It is useful to 
introduce the dimensionless resistance $\rho = R/R_{\rm Q}$ where 
$R_{\rm Q}=h/4e^2$ is the resistance quantum for Cooper pairs.

In the classical limit this system can be described by the Langevin
equation \cite{barone,kautz90}
\begin{equation}
C\left(\frac{\hbar\ddot\varphi}{2e}\right)+\frac{1}{R}
\left(\frac{\hbar\dot\varphi}{2e}\right)+I_{\rm
c}\sin(\varphi)=\frac{V_R}{R}.
\label{eq:phasdiff}
\end{equation}
The term on the righthandside represents the Nyquist noise with $\langle
\delta V_{\rm R}(t+\tau)\delta V_{\rm R}(t)\rangle = 
(2R/\beta)\delta(\tau)$ for a resistor at inverse temperature 
$\beta=1/k_{\rm B}T$. Here, $\delta V_{\rm R}$ denotes the fluctuation 
of the voltage about its average value.

\begin{figure}
\begin{center}
\leavevmode
\epsfxsize=15cm
\epsfbox{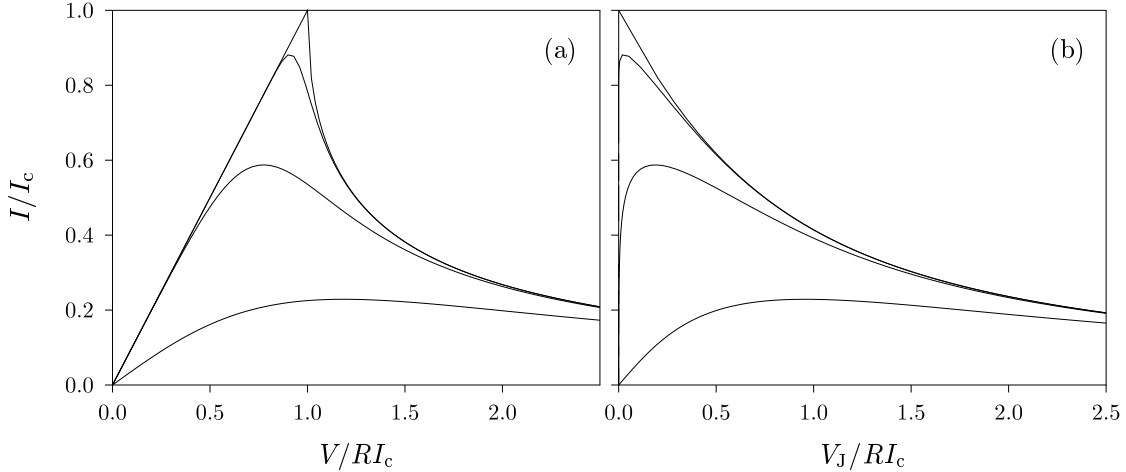}
\end{center}
\caption{(a) Cooper pair current-voltage characteristics $I(V)$
[eq.~(\ref{eq:ivanzilb})] in the overdamped phase diffusion regime 
for $\beta E_{\rm J}= 1, 5, 50,$ and $\infty$ from the lower to the 
upper curve and (b) corresponding characteristics $I(V_{\rm J})$.}
\label{fig:ivan}
\end{figure}

This problem has been solved by Ivanchenko and Zil'berman \cite{ivanc69}
in the overdamped limit where $2\pi^2\rho^2E_{\rm J}\ll E_{\rm c}$. The
stationary solution of the Fokker-Planck equation corresponding to 
eq.~(\ref{eq:phasdiff}) can be expressed in terms of modified Bessel
functions of complex order and yields the Cooper pair current
\cite{ivanc69}
\begin{equation}
\frac{I}{I_{\rm c}}={\rm Im}\left(\frac{I_{1-\ii\beta E_{\rm
J}v}
(\beta E_{\rm J})}{I_{-\ii\beta E_{\rm J}v}(\beta E_{\rm J})}\right).
\label{eq:ivanzilb}
\end{equation}
as a function of the dimensionless applied voltage $v=V/RI_{\rm c}$.
While the supercurrent at zero voltage is destroyed by the fluctuations
causing the phase diffusion, the current-voltage characteristics 
(\ref{eq:ivanzilb}) display peaks at small but finite voltages as 
shown in fig.~\ref{fig:ivan}a. With decreasing temperature the 
$I$--$V$--curve (\ref{eq:ivanzilb}) becomes closer to an ohmic line 
for voltages up to $RI_{\rm c}$, while the peak in the 
$I$--$V_{\rm J}$--curve in fig.~\ref{fig:ivan}b  moves towards zero 
voltage.

We analyze the result (\ref{eq:ivanzilb}) by considering the zero bias
differential resistance 
\begin{equation}
R_0 = \left.\frac{\partial V_{\rm J}}{\partial I}
\right\vert_{V_{\rm J} = 0} =
\left.\frac{\partial V}{\partial I}\right\vert_{V=0} - R.
\label{eq:zbdr}
\end{equation}
$R_0$ is defined with respect to the junction voltage $V_{\rm J}$, and 
$1/R_0$ describes the slope at the origin in fig.~\ref{fig:ivan}b. From 
eq.~(\ref{eq:ivanzilb}) one obtains
\begin{equation}
\frac{R_0}{R} = \frac{1}{I_0^2(\beta E_{\rm J})-1}.
\label{eq:zbdriz}
\end{equation}
For $\beta E_{\rm J}\gg 1$, the $I$--$V$--curve is very
close to $I=V/R$ with an exponentially small difference
\begin{equation}
\frac{R_0}{R} = 2\pi\beta E_{\rm J}\exp(-2\beta E_{\rm J}).
\label{eq:iztherm}
\end{equation}
This shows that for temperatures much lower than the height of the
periodic potential for $\varphi$ in eq.~(\ref{eq:hamil1}), the dynamics 
of the phase is thermally activated.

On the other hand, for $\beta E_{\rm J}\ll 1$ one finds in the
overdamped limit
\begin{equation}
\frac{R_0}{R} = \frac{2}{(\beta E_{\rm J})^2}.
\label{eq:izlow}
\end{equation}
In this case, the $I$--$V$--curve becomes
\begin{equation}
I = \frac{I_{\rm c}^2}{2}\frac{RV}{V^2+(2eR/\hbar\beta)^2} 
\label{eq:ivlow}
\end{equation}
which corresponds to a broad peak structure.

\section{Coulomb Blockade in Josephson Junctions}
\label{sec:coulombblockade}

We now turn to the case of ultrasmall tunnel junctions with spatial
dimensions so small that $E_{\rm c} \gg E_{\rm J}$. Then the phase 
can no longer be treated as a quasiclassical variable, rather the 
charge on the junction capacitance will approximately follow classical 
statistics. As we have seen above, due to phase fluctuations, a 
supercurrent at zero voltage is no longer possible. However, a Cooper 
pair current may flow at finite voltages if the electromagnetic 
environment is able to absorb the energy $2eV$ gained by a Cooper pair 
tunneling through the junction. In contrast to 
sec.~\ref{sec:phasediffusion}, it is now necessary the introduce the 
external impedance on the quantum level. Due to its linearity the
external impedance may be modeled by a possibly infinite number 
of $LC$-oscillators. The corresponding Hamiltonian then reads
\begin{equation}
H_{\rm imp} = \sum_{n=1}^{\infty}\left[\frac{q_n^2}{2C_n}+
\left(\frac{\hbar}{2e}\right)^2\frac{1}{2L_n}(\varphi_{\rm R}-
\varphi_n)^2\right].
\label{eq:imphamil}
\end{equation}
The external impedance in terms of the inductances $L_n$ and
capacitances $C_n$ is given by
\begin{equation}
Z(\omega) = \left[\int_0^{\infty}\dd t \ee^{-\ii\omega t}
\sum_{n=1}^{\infty}
\frac{\cos(\omega_n t)}{L_n}\right]^{-1}
\label{eq:zomeg}
\end{equation}
where $\omega_n = (L_nC_n)^{-1/2}$.  The coupling between the external 
impedance and the phase difference $\varphi$ appears through the phase 
\begin{equation}
\varphi_{\rm R} = \frac{2e}{\hbar}\int_0^t \dd t'(V-V_{\rm J}) =
\frac{2e}{\hbar}
Vt-\varphi.
\label{eq:phir}
\end{equation}
This phase may formally be attributed to the resistance by making use of
the Josephson relation (\ref{eq:josrelv}).

Calculating the Cooper pair current from the Hamiltonian
\begin{equation}
H = H_{\rm J} + H_{\rm imp}
\label{eq:htot}
\end{equation}
perturbatively to lowest order in $E_{\rm J}$, one obtains
\cite{averi90,ingol91}
\begin{equation}
I = \frac{\pi e E_{\rm J}^2}{\hbar}\left[P(2eV)-P(-2eV)\right].
\label{eq:ivpe}
\end{equation}
Here, $P(2eV)$ and $P(-2eV)$ are the probabilities that the energy 
$2eV$ of the tunneling Cooper pair is absorbed or provided by the 
environment, respectively. This probability depends on the external 
impedance and on temperature through
\cite{devor90}
\begin{equation}
P(E) = \frac{1}{2\pi\hbar}\int_{-\infty}^{+\infty}\dd t\exp\left[J(t)+
\frac{\ii}{\hbar}Et\right]
\label{eq:pe}
\end{equation}
with
\begin{equation}
J(t) = 2\int_{-\infty}^{\infty}\frac{\dd\omega}{\omega}\frac{{\rm
Re}Z_{\rm t}
(\omega)}{R_Q}\frac{\ee^{-\ii\omega t}-1}{1-\ee^{-\beta\hbar\omega}}.
\label{eq:jt}
\end{equation}
This expression contains the total impedance seen by the Josephson
junction
\begin{equation}
Z_{\rm t}(\omega) = \frac{1}{\ii\omega C + Z^{-1}(\omega)}
\label{eq:totimp}
\end{equation}
which is given by the external impedance in parallel with the
capacitance.

\begin{figure}
\begin{center}
\leavevmode
\epsfxsize=9.5cm
\epsfbox{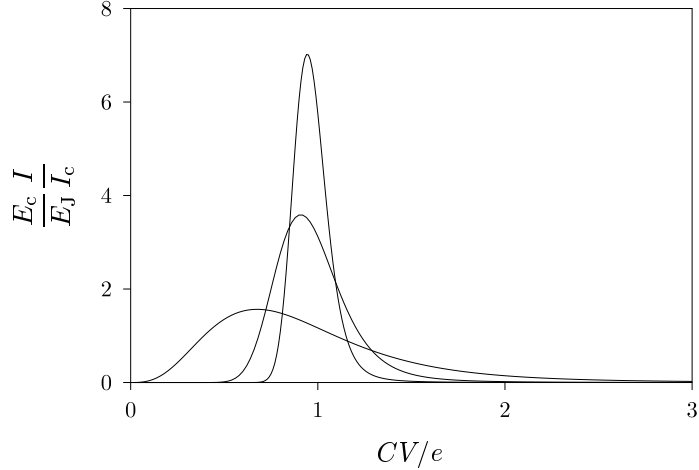}
\end{center}
\caption{Cooper pair current-voltage characteristics $I(V)$ in the
Coulomb blockade regime at zero temperature. The peak is sharpening 
with increasing $\rho = 2, 20, 100$.}
\label{fig:coulombpeak}
\end{figure}

For an ohmic environment [$Z(\omega) = R = \rho R_{\rm Q}$] 
the result (\ref{eq:ivpe}) may be evaluated further.
At zero temperature one finds a zero bias anomaly given by 
$I \sim V^{2\rho -1}$. This shows that at $T=0$ the perturbative
result (\ref{eq:ivpe}) is only valid for large $\rho > 1/2$ where
the current remains small for all voltages. In this case the
$I$--$V$--curve displays a peak near voltages of order $e/C$ as shown 
in fig.~\ref{fig:coulombpeak}. This can easily be understood from a 
simple Coulomb blockade picture. At $V=e/C$ the voltage gain $2eV$ 
just equals the charging energy $E_{\rm c}$.

However, the experimentally relevant situation is an environment  
of low impedance ($\rho\ll1$) \cite{devor91}. On the one hand, it is 
difficult to fabricate large resistances at frequencies around
$E_c/\hbar$, which typically is of the order of 10GHz, and on the 
other hand the resistance should not be too large to avoid heating.
Furthermore, since charging effects are washed out by thermal
fluctuations, we are interested in the regime of low but finite 
temperatures $\beta E_{\rm c} \gg1$. To proceed for parameters in 
this range, we first note that the function $J(t)$ is equivalent 
to the position autocorrelation function of a free damped particle 
and can be evaluated in closed form for ohmic damping \cite{grabe88}. 
For long times, $J(t)$ describes diffusive behavior linear in
time for finite temperatures and logarithmic in time at zero
temperature.  We therefore neglect exponentially decaying terms. This 
restricts us to low voltages which is the regime of interest. 
Assuming furthermore $\rho\ll\beta E_{\rm c}$ in accordance with the 
above considerations, we arrive at \cite{ingol94b}
\begin{equation}
J(t) = -2\rho\left(\ln\left[\frac{\beta E_c}{\pi^2\rho}\sinh\left(
\frac{\pi t}{\hbar\beta}\right)\right] + \gamma +
\ii\frac{\pi}{2}{\rm sign}(t)\right)
\label{eq:jtudo}
\end{equation}
Here, $\gamma = 0.5772\dots$ is the Euler constant. Inserting
(\ref{eq:jtudo}) into (\ref{eq:pe}) and making use of (\ref{eq:ivpe})
one
finally obtains \cite{ingol94b,grabe93,ingol94a}
\begin{equation}
I = {\pi\over 2}I_c{E_J\over E_c}\rho^{2\rho}\left({\beta E_c
\over 2\pi^2}\right)^{1-2\rho}\exp[-2\gamma\rho] {\vert\Gamma(\rho-
\ii{\displaystyle{\beta eV\over \pi}})\vert^2\over \Gamma(2\rho)}
\sinh(\beta eV).
\label{eq:ivudo}
\end{equation}
This perturbative result is only correct if the current is not too 
large which may be the case for very small temperatures and very small
damping. In fact, as noted above, at zero temperature one finds a 
divergence at low voltages. Therefore, the validity is restricted to 
not too low temperatures $\beta E_{\rm J} \ll\rho$. Since 
$E_{\rm c}\gg E_{\rm J}$, this together with the upper bound on 
temperature [$\beta E_{\rm c}\gg\rho$] still leaves a rather large 
range where (\ref{eq:ivudo}) is applicable.

The current-voltage characteristics (\ref{eq:ivudo}) is shown in
fig.~\ref{fig:ivs} (see sec.~\ref{sec:between}) as dashed line for 
$\beta E_{\rm J}=2, \rho=0.04$ and $\beta E_{\rm c}= 1$ and $10^5$. 
It exhibits a peak at finite voltage which for small $\rho$ is given by
\begin{equation}
V_{\rm max} = \frac{\pi\rho}{e\beta}(1+4\zeta(3)\rho^3+\dots).
\label{eq:vmax}
\end{equation}
Here $\zeta(3)=1.202\dots$ is a Riemann number. In the limit of small
$\rho$, high temperatures and for $\beta eV\ll1$ one recovers the result
of classical phase diffusion (\ref{eq:ivlow}).

A {\sl nonohmic environment} of practical interest is a finite $LC$
transmission line terminated by an ohmic load resistance $R_L$
\cite{holst94,holst94a}. The transmission line is characterized by its 
resistance at infinite length $R_\infty = (L_0/C_0)^{1/2}$ where $L_0$ 
and $C_0$ are specific inductance and capacitance per unit length. As 
a parameter describing the finite length $\ell$ of the transmission line 
we choose the $\lambda/4$-frequency $\omega_0 = (\pi/2)u/\ell$ where 
$u=(L_0C_0)^{-1/2}$ is the velocity of wave propagation on the line. 

\begin{figure}
\begin{center}
\leavevmode
\epsfxsize=9.6cm
\epsfbox{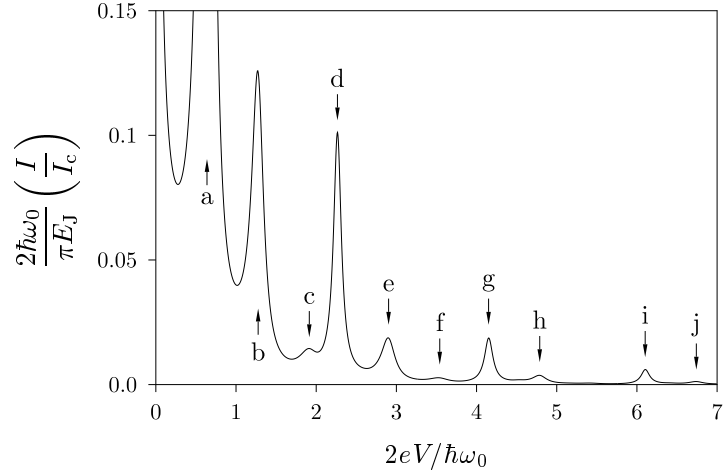}
\end{center}
\caption{Cooper pair current-voltage characteristic for an ultrasmall
tunnel junction coupled to a finite transmission line with $R_{\rm L}/
R_{\rm Q}=0.01, R_{\rm L}/R_\infty=0.1,$ and $\omega_0R_\infty C=1$ 
at zero temperature. The peaks correspond to excitations denoted by 
$(N_1N_2N_3N_4)$, where $N_k$ is the number of quanta of the $k$-th mode 
excited. a: (1000), b: (2000), c: (3000), d: (0100), e: (1100), f:
(2100), g: (0010), h: (1010), i: (0001), j: (1001).}
\label{fig:resiv}
\end{figure}

Depending on the ratio between load resistance and resistance of the
infinite transmission line, the environment behaves quite differently. 
For $R_L = R_\infty$ the impedance matching results in an ohmic 
impedance. On the other hand, for $R_L$ very different from $R_\infty$, 
the impedance displays sharp resonances. According to the above 
discussion, the properties of the external impedance should show up in 
the probability to absorb or emit energy, i.e.\ in $P(E)$, and therefore 
according to (\ref{eq:ivpe}) in the Cooper pair current-voltage 
characteristics. This is indeed the case as can be seen from
fig.~\ref{fig:resiv}. There we show the zero temperature Cooper pair
current for $R_L/R_\infty=0.1$ obtained from a direct numerical 
evaluation of eqs.\ (\ref{eq:ivpe})--(\ref{eq:jt}).
The resonance peaks can be identified as single or multiple excitations
of various transmission line modes \cite{ingol94b}. The measurement of 
the Cooper pair current thus allows for a spectroscopy of the 
environmental modes.

\section{Supercurrent Peak: Between the Phase Diffusion and Coulomb
Blockade
Regimes}
\label{sec:between}

So far, we have discussed the behavior of a Josephson junction in the
classical overdamped limit (sec.~\ref{sec:phasediffusion}) and in the
quantum regime for charging energies much larger than the Josephson 
coupling energy (sec.~\ref{sec:coulombblockade}). We now want to bridge 
the gap between these two limits by extending the perturbation theory 
presented in the previous section to infinite order in $E_{\rm J}$. 

The summation of the perturbation series in $E_{\rm J}$ is only possible
for an appropriate choice of the parameter regime. Motivated by the above 
discussion, we will continue to consider the overdamped regime where 
$\omega_R=1/RC$ is much larger than the Josephson frequency 
$\omega_{\rm J}=(2e/\hbar) RI_{\rm c}$. This is identical to the
condition $2\pi^2 \rho^2 E_{\rm J} \ll E_{\rm c}$ given in 
sec.~\ref{sec:phasediffusion}. In addition, the correlation function 
$J(t)$ contains a thermal frequency scale given by $\nu=2\pi/\hbar\beta$
as can be seen from the last denominator in eq.~(\ref{eq:jt}). 
In the following we assume that $\omega_{\rm J}\ll\nu$ so that we may 
neglect terms exponentially decaying in time with $\nu$ or faster. This 
results in the approximation 
\begin{equation}
J(t) = -2\rho\left[\frac{\pi}{\hbar\beta}\vert t\vert + S +
\ii\frac{\pi}{2}
{\rm sign}(t)\right]
\label{eq:jt1}
\end{equation}
valid for sufficiently high temperatures where $\rho\beta E_{\rm J}\ll 
1$. Further, we have introduced the abbreviation
\begin{equation}
S=\gamma+\frac{\pi^2\rho}{\beta E_{\rm c}}+\psi\left(\frac{\beta E_{\rm
c}}
{2\pi^2\rho}\right)
\label{eq:s}
\end{equation}
where $\psi(x)$ denotes the logarithmic derivative of the gamma
function.

For the correlation function (\ref{eq:jt1}) the Cooper pair current can
be evaluated exactly and expressed in terms of a continued fraction
\cite{grabe98}
\begin{equation}
I = I_{\rm c}{\rm Re}\left[\frac{\sin(\pi\rho)}
{2\pi\rho}\frac{\exp(-2\rho S)}{v+\ii/\beta E_{\rm J}}
\frac{\displaystyle 1}{\displaystyle 1+\frac{\displaystyle b_1}
{\displaystyle 1+\frac{\displaystyle b_2}{\displaystyle
1+\dots}}}\right] 
\label{eq:is3}
\end{equation}
with coefficients
\begin{equation}
b_n=\left(\frac{\beta E_{\rm J}}{2\pi\rho}\right)^2\frac{\sin(\pi\rho n)
\sin(\pi\rho(n+1)) \exp(-2\rho S)}
{n(n+1)(n-\ii v\beta E_{\rm J})(n+1-\ii v\beta E_{\rm J})}.
\label{eq:cfcoeff}
\end{equation}
Since the continued fraction converges rapidly, we may linearize the
sine functions appearing in eq.~(\ref{eq:cfcoeff}) for sufficiently 
small $\rho$. Then, the continued fraction may be evaluated with the 
help of a matrix recursion and one finds \cite{grabe98}
\begin{equation}
I=\frac{2e}{\hbar}E_{\rm J}^*\,{\rm Im}\left(\frac{I_{1-\ii\beta
eV/
\pi\rho}(\beta E_{\rm J}^*)}{I_{-\ii\beta eV/\pi\rho}(\beta E_{\rm
J}^*)}\right).
\label{eq:genivanzilb}
\end{equation}
with an effective Josephson energy
\begin{equation}
E_{\rm J}^*=E_{\rm J}\exp(-\rho[\psi(1+\hbar\beta\omega_{\rm R}
/2\pi)+\gamma]).
\label{eq:ejeff}
\end{equation}
Here we have rewritten the quantity $S$ defined in eq.~(\ref{eq:s}) 
in terms of the frequency $\omega_R$ and have neglected a term
$\pi\rho/\hbar\beta\omega_R\ll1$ in the overdamped limit considered. 
Even though $\rho$ has to be small for the expression (\ref{eq:ejeff}) 
to hold, the correction to the bare Josephson energy may be substantial 
because for low temperatures the $\psi$ function grows logarithmically 
and may become large. On the other hand, in the classical limit 
where $\hbar\to 0$ and $e\to 0$ such that the flux quantum $h/2e$ 
remains constant, the effective Josephson energy $E_{\rm J}^*$ coincides 
with the bare Josephson energy $E_{\rm J}$. We thus recover in the 
classical overdamped limit the result (\ref{eq:ivanzilb}).

\begin{figure}
\begin{center}
\leavevmode
\epsfxsize=9.1cm
\epsfbox{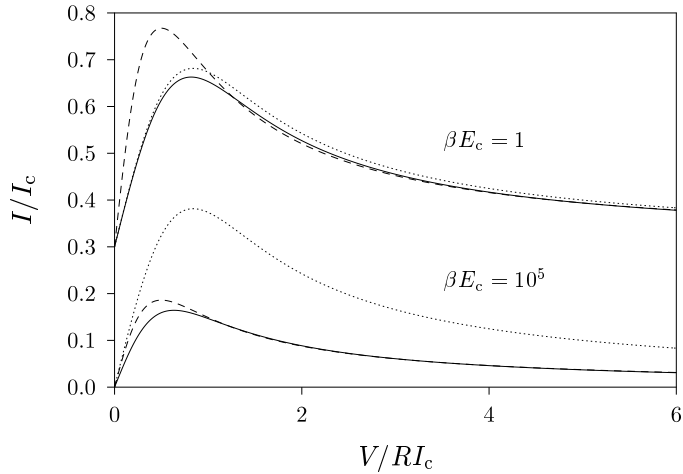}
\end{center}
\caption{The current-voltage characteristic of a Josephson junction
with Josephson energy $\beta E_{\rm J}=2$ is shown for charging energies
$\beta E_{\rm c}=1$ and $10^5$ and external resistance $\rho=0.04$. The
full line corresponds to our result (\protect\ref{eq:is3}) 
while the dotted line gives the standard Ivanchenko-Zil'berman 
(\ref{eq:ivanzilb}) result and the dashed line depicts the prediction 
(\protect{\ref{eq:ivudo}}) for Coulomb blockade. The two sets of 
current-voltage characteristics are vertically shifted with respect to 
each other by $I/I_{\rm c}=0.3$ for sake of clarity.}
\label{fig:ivs}
\end{figure}

In fig.~\ref{fig:ivs} we show how the result (\ref{eq:genivanzilb})
bridges between the phase diffusion result (\ref{eq:ivanzilb}) shown as 
dotted line and the Coulomb blockade result (\ref{eq:ivudo}) depicted 
as dashed line. The current-voltage characteristics have been calculated 
for $\beta E_{\rm J} = 2$ and $\rho=0.04$. For large $\beta E_{\rm c}$ 
where charging effects should be important one obtains very good 
agreement with the Coulomb blockade result. With decreasing 
$\beta E_{\rm c}$ the current-voltage characteristics evolve 
differently and a crossover to the Ivanchenko-Zil'berman result
is found.

\section{Conclusions}
\label{sec:conclude}
We have studied  the current-voltage characteristics of mesoscopic
Josephson junctions focusing on the  modifications of the
supercurrent as the junction parameters change. We have not
discussed here the effect of the charging energy on quasiparticle
tunneling relevant at higher applied voltages only \cite{falci}.
As the charging energy $E_{\rm c}$ grows relative to the Josephson 
energy $E_{\rm J}$, the supercurrent of the classical Josephson effect 
was shown to evolve gradually into a supercurrent peak caused by 
incoherent Cooper pair tunneling. In the Coulomb blockade regime two 
types of structures may appear for small junctions with $E_J\ll E_c$ 
embedded in a standard low impedance environment. The first structure, 
a peak at low voltages has recently been seen in experiments on 
lithographically fabricated junctions \cite{holst94,holst94a} as well 
as break junctions \cite{mulle94}. The second  structure, peaks in the
current-voltage characteristics due to resonances in the environmental
impedance have also been seen in experiments \cite{holst94,holst94a}
with a well-defined environment consisting of two transmission
line segments allowing for a quantitative test of the theoretical 
predictions and good agreement was found. Although detailed experimental 
studies of the region between the phase diffusion and Coulomb blockade 
limits  are absent, recent work \cite{vion96} indicates quantum effects 
in qualitative accord with the predictions made.

We would like to thank M.~H.~Devoret,  D.~Esteve, and C.~Urbina for 
stimulating discussions. One of us (HG) was supported by
the Deutsche Akademischer Austauschdienst (DAAD), while support for
the other author (GLI) during a stay at the Centre d'Etudes de Saclay 
was provided by the Volkswagenstiftung.

\end{document}